\documentstyle[12pt]{article}
\begin{document}
\centerline{\Large\bf Homogeneous and isotropic cosmologies}
\vskip .1in
\centerline{\Large\bf with nonlinear electromagnetic radiation}
\vskip .6in
\centerline{Dan N. Vollick}
\vskip .1in
\centerline{Irving K. Barber School of Arts and Sciences}
\centerline{University of British Columbia Okanagan}
\centerline{3333 University Way}
\centerline{Kelowna, B.C.}
\centerline{Canada}
\centerline{V1V 1V7}
\vskip .1in
\centerline{and}
\vskip .1in
\centerline{Pacific Institute for Theoretical Physics}
\centerline{Department of Physics and Astronomy}
\centerline{University of British Columbia}
\centerline{6224 Agricultural Road}
\centerline{Vancouver, B.C.}
\centerline{Canada}
\centerline{V6T 1Z1}
\vskip .8in
\centerline{\bf\large Abstract}
\vskip 0.4in
\noindent
In this paper I examine cosmological models that contain a stochastic background of nonlinear electromagnetic radiation.
I show that for Born-Infeld electrodynamics the equation of state parameter, $w=P/\rho$, remains close to 1/3 throughout
the evolution of the universe if $E^2=B^2$ in the late universe to a high degree of accuracy.

Theories with electromagnetic Lagrangians of the form
$L=-\frac{1}{4}F^2+\alpha F^4$
have recently been studied in magnetic universes, where the electric field vanishes. It was shown that the $F^4$ term can
produce a bounce in the early universe, avoiding an initial singularity. Here I show that the inclusion of an electric
field, with $E^2\simeq B^2$ in the late universe, eliminates the bounce and the universe ``begins" in an initial singularity.

I also examine theories with Lagrangians of the form
$L=-\frac{1}{4}F^2-\mu^8/F^2$,
which have been shown to produce a period of late time accelerated expansion in magnetic universes. I show that, if an electric field
is introduced, the accelerated phase will only occur if $E^2<3B^2$.
\newpage
\section{Introduction}
Over the last few years there has been a significant amount of interest in cosmological models involving
nonlinear electromagnetic fields \cite{No1,No2,BI}. If the
early universe is dominated by radiation governed by Maxwell's equations it is well known that there will be a
spacelike initial singularity in the past. However, if Maxwell's equations become modified in the early universe, when the
electromagnetic field is large, it may be possible to avoid the initial singularity. In fact, recent results \cite {No1} show that
a magnet universe can avoid the initial singularity and have a period of late time acceleration if the electromagnetic Lagrangian is of the form
\begin{equation}
L=-\frac{1}{4}F^2+\alpha F^4-\frac{\mu^8}{F^2} ,
\label{eq1}
\end{equation}
where $F^2=F^{\mu\nu}F_{\mu\nu}$ and $\alpha$ and $\mu$ are constants. In these models the universe begins in a collapsing phase
and the scale factor decreases until it reaches some minimum value and the universe then begins to expand. Near the bounce the electromagnetic
filed is large and the $\alpha F^4$ term in the Lagrangian dominates. At late times the $\mu/F^2$ term dominates and the universe
enters an accelerated expansion phase. It has also been shown \cite{No2} that if a term proportional to $1/F^4$ is added to the Lagrangian
the expansion will eventually end and the universe will begin to collapse, until it bounces again. Thus, this model produces a cyclic magnetic
universe.

In this paper I examine cosmological spacetimes that contain a stochastic background of Born-Infeld radiation
with a non-vanishing $<E^2>$. The equation of state parameter, $w=P/\rho$,
is computed as a function of the scale factor and is shown to be $w\simeq 1/3$
throughout the history of the universe if $<E^2>\simeq <B^2>$ at late times.
This implies that the nonlinear corrections to Maxwell's
equations, that appear in Born-Infeld theory, do not significantly effect the evolution of the universe.

I also examine the early universe in theories with
\begin{equation}
L=-\frac{1}{4}F^2+\alpha F^4
\end{equation}
which, as discussed above, have a bounce at a small value of the scale factor in magnetic universes. Here I show that the
inclusion of a stochastic electric field keeps the $\alpha F^4$ term small in comparison to the $F^2$ term, if $<E^2>\simeq <B^2>$
at late times. Thus, the inclusion of an electric field can
eliminate the bounce and the universe can ``begin" from an initial singularity.

I also showed that terms proportional to $1/F^2$ in the Lagrangian, which dominate at late times, will only produce an accelerated expansion
if $E^2<3B^2$.
\section{Nonlinear Electrodynamics}
In nonlinear electrodynamics the Maxwell Lagrangian
\begin{equation}
L_M=-\frac{1}{4}F^2
\end{equation}
is replaced by
\begin{equation}
L=L(F^2,G^2)\; ,
\end{equation}
where $F^2=F^{\mu\nu}F_{\mu\nu}$, $G^2=F^{*\mu\nu}F_{\mu\nu}$ and $F^{*\mu\nu}$ is the dual
of $F_{\mu\nu}$. In this paper I will take $L$ to be independent of $G$ for simplicity.

The vacuum field equations of the theory are
\begin{equation}
\nabla_{\mu}P^{\mu\nu}=0
\end{equation}
and
\begin{equation}
\nabla_{\mu}F^{*\mu\nu}=0
\end{equation}
where
\begin{equation}
P^{\mu\nu}=\frac{\partial L}{\partial F_{\mu\nu}}\;.
\end{equation}
In a cosmological spacetime with
\begin{equation}
ds^2=-dt^2+a(t)^2\left[dx^2+dy^2+dz^2\right]\;.
\label{metric}
\end{equation}
the field equations can be written as
\begin{equation}
\vec{\nabla}\cdot\vec{D}=0,\;\;\;\;\;\;\;\;\;\;\;\;\;\;\;\;\;\;
\frac{\partial}{\partial t}\left( a^2\vec{D}\right)-a\vec{\nabla}\times\vec{H}=0\; ,
\label{E}
\end{equation}
\begin{equation}
\vec{\nabla}\cdot\vec{B}=0,\;\;\;\;\;\;\;\;\;\;\;\;\;\;\;\;\;\; \frac{\partial}{\partial t}\left( a^2\vec{B}\right)+
a\vec{\nabla}\times\vec{E}=0\; ,
\label{B}
\end{equation}
where $E_k=a^{-1}F_{kt}$, $B_k=\frac{1}{2}a^{-2}\epsilon_{klm}F_{lm}$,
$\vec{D}=-4L_F\vec{E}$, $\vec{H}=-4L_F\vec{B}$, $F^2=2(B^2-E^2)$
and $L_F=dL/dF^2$. The factors of $-4$ appear so that $\vec{D}\simeq\vec{E}$ and $\vec{H}\simeq\vec{B}$
in the weak filed limit where
$L\simeq -\frac{1}{4}F^2$.
The energy-momentum tensor can be found by varying the action with respect to the metric and is given by
\begin{equation}
T^{\mu\nu}=-2P^{\mu\alpha}F^{\nu}_{\;\;\;\alpha}+g^{\mu\nu}L
\label{EM}
\end{equation}
Born and Infeld \cite{Bo1} took the Lagrangian to be (setting $G=0$)
\begin{equation}
L=-\frac{1}{2b^2}\left[\sqrt{1+b^2F^2}-1\right]\;.
\label{BI}
\end{equation}
It is interesting to note that the action for gauge theories on D-branes in string theory is of the Born-Infeld
type \cite{Po1}.
\section{Cosmologies with Nonlinear Electromagnetic Radiation}
In this section I will consider $k=0$ homogeneous and isotropic cosmological spacetimes, with a metric given
in (\ref{metric}), that is filled with electromagnetic radiation.
The electromagnetic field that is of cosmological interest is the cosmic microwave background. It can be
considered as a stochastic background of short wavelength radiation (compared to the curvature) that satisfies \cite{To1}
\begin{equation}
<E_i>=<B_i>=<E_iB_j>=0
\end{equation}
and
\begin{equation}
<E_iE_j>=\frac{1}{3}E^2\delta_{ij}\;\;\;\;\;\;\;\;\;\;\;\;
<B_iB_j>=\frac{1}{3}B^2\delta_{ij}\; ,
\end{equation}
where $<\;>$ denotes an average over a volume that is large compared to the wavelength of the radiation
but small compared to the curvature of the spacetime.
I will also assume that approximations such as
$<f(E^2)>\simeq f(<E^2>)$ are valid for functions $f$ that appear in this paper.
This type of approximation is used in many papers that examine cosmologies with nonlinear electromagnetic radiation but,
is often not stated.
In this paper I will take $<E^2>\simeq <B^2>$ at late times, if
Maxwell's equations are approximately valid then.
For notational simplicity I will omit the averaging
brackets for the remainder of this paper.

The energy density and pressure of the radiation can be found from (\ref{EM}) using $\rho=-T^t_{\;\;\; t}$ and
$P=\frac{1}{3}T^k_{\;\;\; k}$ and are given by
\begin{equation}
\rho=-L-4E^2L_F
\end{equation}
and
\begin{equation}
P=L-\frac{4}{3}\left(E^2+F^2\right)L_F\;.
\end{equation}

The behavior of $D^2$ and $B^2$ can be found by multiplying the second equation in (\ref{E}) by
$a^2\vec{D}$ and the second equation in (\ref{B}) by $a^2\vec{B}$ and taking spatial averages to obtain
\begin{equation}
\frac{\partial}{\partial t}\left( a^4D^2\right)=2a^3<\vec{D}\cdot(\vec{\nabla}\times\vec{H})>
\end{equation}
and
\begin{equation}
\frac{\partial}{\partial t}\left( a^4B^2\right)=-2a^3<\vec{B}\cdot(\vec{\nabla}\times \vec{E})>\; .
\end{equation}
Since the right hand sides of these equations vanishes we find that
\begin{equation}
D^2=\frac{D_0^2}{a^4}\;\;\;\;\;\;\;\;\;\; and \;\;\;\;\;\;\;\;\ B^2=\frac{B_0^2}{a^4}\; ,
\label{fields}
\end{equation}
where $D_0$ and $B_0$ are the present values of $D$ and $B$ and $a$ is taken to be one today.
Solving for $E^2$, using the Born-Infeld Lagrangian, gives
\begin{equation}
E^2=\frac{D_0^2}{a^4}\left[\frac{1+\frac{2b^2B_0^2}{a^4}}{1+\frac{2b^2D_0^2}{a^4}}\right]\; .
\end{equation}

The equation of state parameter $w$ defined by
\begin{equation}
w=\frac{P}{\rho}
\end{equation}
is given by
\begin{equation}
w=\frac{1}{3}-\frac{4(L-F^2L_F)}{3(L+4E^2L_F)}\;.
\end{equation}
For theories that reduce to Maxwell's theory in the weak field limit
\begin{equation}
L\simeq -\frac{1}{4}F^2\left[ 1+(\alpha F^2)^n\right]\;\;\;\;\;\;
\;\;\;\;\; n>0\;,
\end{equation}
when $\alpha E^2<<1$ and $\alpha B^2<<1$. The equation of state parameter is
given by
\begin{equation}
w\simeq \frac{1}{3}+\frac{2n(\alpha F^2)^{n+1}}{3\alpha(E^2+B^2)}\;,
\end{equation}
so that $w\simeq 1/3$, as expected.
For Born-Infeld theory the equation of state paramter is given by
\begin{equation}
w(a)=\frac{1}{3}-\frac{\frac{4}{3}a^4\left[1-\frac{a^4+b^2(D_0^2+B_0^2)}
{\sqrt{(a^4+2b^2D_0^2)(a^4+2b^2B_0^2)}}\right]}{a^4-\sqrt{(a^4+2b^2D_0^2)(a^4+2b^2B_0^2)}}\;.
\end{equation}
In the early universe $a\rightarrow 0$ and it is easy to see that $w\rightarrow\frac{1}{3}$.
 Note: if $B_0^2=D_0^2$ then $w=1/3$ at all times. This makes sense because $B_0^2
=D_0^2$ implies that $F^2=0$. At late times $D_0^2\simeq E_0^2\simeq B_0^2$.
Setting $B_0^2=D_0^2+\epsilon$, with $\epsilon<<D_0^2$, we find that
\begin{equation}
w\simeq \frac{1}{3}-\frac{a^4b^2\epsilon^2}{3D_0^2(a^4+2b^2D_0^2)^2}
\end{equation}
to lowest order in $\epsilon$. The equation of state parameter therefore decreases from
$w=\frac{1}{3}$ at $a=0$ to a minimum value of $w_{min}\simeq\frac{1}{3}-\frac{1}{24}(\frac{\epsilon}{D_0^2})^2$
at $a^4=2b^2D_0^2$ and then increases to
$w\simeq\frac{1}{3}-\frac{1}{3}(\frac{b\epsilon}{D_0})^2$ at $a=1$. This shows that $w\simeq 1/3$ at all times.
This can also be seen from
\begin{equation}
b^2F^2=\frac{2b^2\epsilon}{a^4+2b^2D_0^2}\; ,
\end{equation}
which is always small. Thus, Maxwell's equations will hold to a good approximation at all times and $w\simeq 1/3$
(it is easy to show that $\rho+3P\geq 0$ in Born-Infeld theory
for all values of $E^2$ and $B^2$ so that a bounce cannot occur).

An interseting way of looking at the radiation is to consider it to be composed of two interacting
fluids with
\begin{equation}
\rho_1=-4E^2L_F\;\;\;\;\;\;\;\;\;\;\;\;\;\;\;\;\;\;\;\; P_1=-\frac{4}{3}E^2L_F
\end{equation}
and
\begin{equation}
\rho_2=-L\;\;\;\;\;\;\;\;\;\;\;\;\;\;\;\;\;\;\;\;\;\;\;\;\; P_2=L-\frac{4}{3}F^2L_F\;.
\label{fluid2}
\end{equation}
The equation of state for fluid one is $P_1=\frac{1}{3}\rho_1$, so that it behaves like Maxwell radiation (note
that for Maxwell's theory and for Born-Infeld theory $L_F<0$ so that $\rho_1>0$). The equation of state for the second
fluid depends on the form of the Lagrangian. If the fluids are noninteracting the energy density of the first fluid
will satisfy $\rho_1\propto a^{-4}$. However,
\begin{equation}
\rho_1=-\frac{D^2}{4L_F}=-\frac{D_0^2}{4a^4L_F}\; ,
\end{equation}
so that the fluids are noninteracting iff $L_F$ is a constant. If $L_F$ varies by only a small amount then very
little energy will be transferred from one fluid to the other. In Born-Infeld theory $L_F\simeq -1/4$ since
$b^2F^2$ is always small. Thus, the two fluids are almost noninteracting. The energy densities of the fluids are given by
\begin{equation}
\rho_1\simeq E^2\simeq \frac{D_0^2}{a^4}
\end{equation}
and
\begin{equation}
\rho_2\simeq\frac{\epsilon}{2(a^4+2b^2D_0^2)}\; .
\end{equation}
Thus, $|\rho_2|<<\rho_1$ and $w\simeq 1/3$ at all times.

The equation of state for the second fluid can be found using equation (\ref{fluid2}).
For the Born-Infeld Lagrangian (\ref{BI}) the equation of state is given by
\begin{equation}
P_2=\frac{1}{3}\rho_2\left[\frac{1-2b^2\rho_2}{1+2b^2\rho_2}\right]\;.
\end{equation}
At low densities ($|\rho_2|<<b^{-2}$) the equation of state is $P_2\simeq\frac{1}{3}\rho_2$ and at high
densities ($\rho_2>>b^{-2}$) the equation of state is $P_2\simeq-\frac{1}{3}\rho_2$. The minimum value of
$\rho_2$ is $-\frac{1}{2b^2}$, and the pressure diverges as $\rho_2$ approaches this value. Since $b^2\rho_2<<1$
we have $P_2\simeq\frac{1}{3}\rho_2$ during the evolution of the universe.

Next consider the Lagrangian
\begin{equation}
L=-\frac{1}{4}F^2+\alpha F^4
\end{equation}
which, in magnetic universes, does not have an initial singularity for $\alpha >0$. The singularity avoiding behavior is produced by the
$F^4$ term, which dominates at early times. However, in universes with $E^2\neq 0$ it is not necessary that $F^2=2(B^2-E^2)$ is large
in the early universe. From
$F^2=2(B^2-E^2)$ and $\vec{D}=-4L_F\vec{E}$ we find the following cubic equation
\begin{equation}
x^3+(2\lambda B^2-1)x^2-2\lambda D^2=0\; ,
\label{cubic}
\end{equation}
where $x=1-\lambda F^2$ and $\lambda=8\alpha$.
At late times Maxwell's equations will hold to a good approximation and $F^2\simeq 0$. To see how $x$ evolves
set $x=1+\epsilon$, linearize (\ref{cubic}) and solve for $\epsilon$. Using (\ref{fields}) it is easy to show that
\begin{equation}
\epsilon=\frac{2\lambda (D_0^2-B_0^2)}{(a^4+4\lambda B_0^2)}\;,
\end{equation}
Thus, the maximum value of $\epsilon$ is
\begin{equation}
\epsilon_{\max}=\frac{(D_0^2-B_0^2)}{2B_0^2}\;.
\end{equation}
Since $D_0^2-B_0^2<<B_0^2$ we see that $\epsilon_{max}<<1$. Thus, the $\alpha F^4$ in the Lagrangian will never dominate over
the $F^2$ term in the Lagrangian and there will be an initial singularity.

Finally, consider situations in which the $\mu^8/F^2$ term in (\ref{eq1}) dominates. A necessary and sufficient condition for
accelerated expansion is $\rho+3P<0$ (it is simpler to examine $\rho+3P$ than $w$ since the energy density can be negative in
this theory). Now
\begin{equation}
\rho+3P\simeq-\frac{4\mu^8}{F^4}(3B^2-E^2)\;.
\end{equation}
Thus, accelerated expansion will occur iff $E^2<3B^2$
(note that we cannot have $E^2=B^2$, since the energy density would diverge).
\section*{Conclusion}
In this paper I examined homogeneous and isotropic cosmologies with nonlinear electromagnetic radiation. The electromagnetic field was taken to be a
stochastic backgound with non-vanishing $E^2$ and $B^2$. I showed that, for Born-Infeld theory,
the equation of state parameter $w=P/\rho$ is always
close to 1/3 if $E^2$ and $B^2$ are nearly the same in the late universe.

I also examined cosmologies with electromagnetic Lagrangians given by
\begin{equation}
L=-\frac{1}{4}F^2+\alpha F^4
\label{lag1}
\end{equation}
and by
\begin{equation}
L=-\frac{1}{4}F^2-\frac{\mu^8}{F^2}\; .
\label{lag2}
\end{equation}
In magnetic universes with the Lagrangian (\ref{lag1}) the $\alpha F^4$ term dominates in the early universe producing a bounce.
However, the inclusion of an electric field, with $E^2\simeq B^2$ at late times, keeps $\alpha F^2$ small and these models do not have a bounce in the early universe.
At late times in magnetic universes with the Lagrangian (\ref{lag2})
the $\mu^8/F^2$ term will dominate producing an accelerated expansion. I showed that the universe will only experience a period of late time
acceleration if $E^2<3B^2$.

\section*{Acknowledgements}
This research was supported by the  Natural Sciences and Engineering Research
Council of Canada.

\end{document}